\documentclass[11pt]{article}

\usepackage{a4wide}
\usepackage{graphicx}

\newcommand{\dd}{{\rm d}}

\newcommand{\dr}{\partial}

\pdfoutput=1

\begin{document}

\begin{center}   
\textbf{\LARGE Direct numerical simulations of aeolian sand ripples}

\vspace*{0.2cm}

O. \textsc{Dur\'an}$^{\flat,\sharp}$, P. \textsc{Claudin}$^\sharp$ and B. \textsc{Andreotti}$^\sharp$
\end{center}

{\small
\noindent
$^\flat$ MARUM--Center for Marine Environmental Sciences, University of Bremen, Leobener Strasse, D-28359 Bremen, Germany.\\
$^\sharp$ Laboratoire de Physique et M\'ecanique des Milieux H\'et\'erog\`enes (PMMH),
UMR 7636 CNRS -- ESPCI -- Univ. Paris Diderot -- Univ. P.M. Curie, 10 rue Vauquelin, 75005 Paris, France.
}

\begin{abstract}
Aeolian sand beds exhibit regular patterns of ripples resulting from the interaction between topography and sediment transport. Their characteristics have been so far related to reptation transport caused by the impacts on the ground of grains entrained by the wind into saltation. By means of direct numerical simulations of grains interacting with a wind flow, we show that the instability turns out to be driven by resonant grain trajectories, whose length is close to a ripple wavelength and whose splash leads to a mass displacement towards the ripple crests. The pattern selection results from a compromise between this destabilizing mechanism and a diffusive downslope transport which stabilizes small wavelengths. The initial wavelength is set by the ratio of the sediment flux and the erosion/deposition rate, a ratio which increases linearly with the wind velocity. We show that this scaling law, in agreement with experiments, originates from an interfacial layer separating the saltation zone from the static sand bed, where momentum transfers are dominated by mid-air collisions. Finally, we provide quantitative support for the use the propagation of these ripples as a proxy for remote measurements of sediment transport.
\end{abstract}

Observers have long recognized that wind ripples \cite{B41,EEW75} do not form via the same dynamical mechanism as dunes \cite{W72}. Current explanations ascribe their emergence to a geometrical effect of solid angle acting on sediment transport. The motion of grains transported in saltation is composed of a series of asymmetric trajectories \cite{A04,UH87,AH88,KR09} during which they are accelerated by the wind. These grains, in turn, decelerate the airflow inside the transport layer \cite{B41,KR09,Cetal09,DCA11,CPH11,KPMK12,LMcKN12}. On hitting the sand bed, they release a splash-like shower of ejected grains that make small hops from the point of impact \cite{B41,RWMcE96,RVB00}. This process is called reptation. Previous wind ripple models assume that saltation is insensitive to the sand bed topography and forms a homogeneous rain of grains approaching the bed at a constant oblique angle \cite{A87,A90,CMRV00,YBP04,P99,HM99}. Upwind-sloping portions of the bed would then be submitted to a higher impacting flux than downslopes \cite{B41}. With a number of ejecta proportional to the number of impacting grains, this effect would produce a screening instability with an emergent wavelength $\lambda$ determined by the typical distance over which ejected grains are transported \cite{A87,A90,CMRV00}, a few grain diameters $d$. However, observed sand ripple wavelengths are about 1000 times larger than $d$, on Earth. The discrepancy is even more pronounced on Mars, where regular ripples are 20 to 40 times larger than those on a typical earth sand dune \cite{Betal12,SFVBO10}. Moreover, the screening scenario predicts a wavelength independent of the wind shear velocity $u_*$, in contradiction with field and wind tunnel measurements that exhibit a linear dependence of $\lambda$ with $u_*$ \cite{S63,SL78,ACP06}.

\section{Model}
\label{Model}

In order to unravel the dynamical mechanisms resolving these discrepancies, we perform direct numerical simulations of a granular bed submitted to a turbulent shear flow. This flow is driven by a turbulent shear stress $\rho_f u_*^2$ imposed far from the bed, where $\rho_f$ denotes air density or more generally that of the fluid constituting the atmosphere. The grains, of density $\rho_p$, are subject to gravity $g$ and interact through contact forces. Based on the work of \cite{DAC12}, we explicitly implement a two-way coupling between a discrete element method for the particles and a continuum Reynolds averaged description of hydrodynamics, coarse-grained at a scale larger than the grain size. This coupling occurs by means of drag and Archimedes forces in the equations of motion of the grains, and via a body force term in the RANS equations ({\it SI text}). This method enables us to perform runs over long periods of times using a large two-dimensional spatial domain, while keeping the whole complexity of the granular phase (Video~S1). In particular, we do not have to introduce a splash function to describe reptation: the ejecta generated by a grain that collides with the sediment bed are directly obtained from the interaction of the particles with their neighbors in contact.

\begin{figure*}
\centerline{\includegraphics{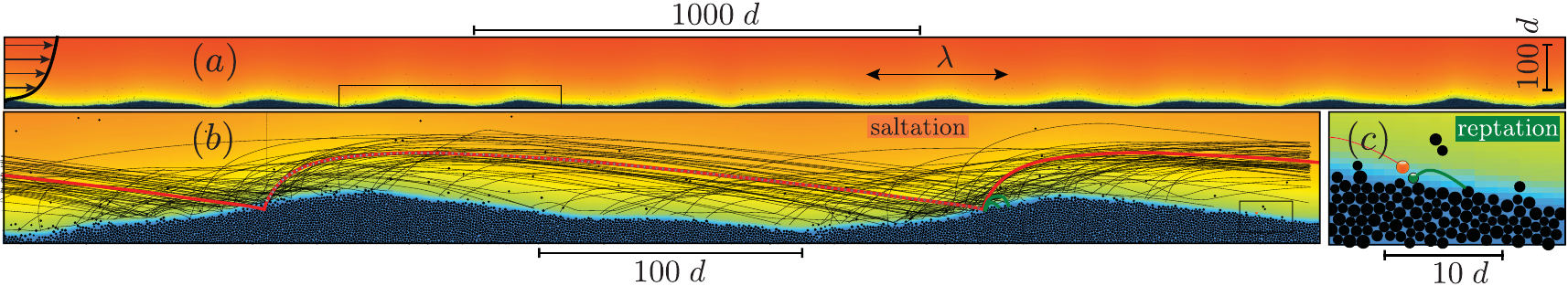}}
\caption{Ripples emerging from a flat bed in a simulation ($u_*/u_{\it th} = 3$). (a) Large-scale view of the system composed of $45000$ grains in a quasi two-dimensional $xyz$ box of respective dimensions $3400\,d \times 1\,d \times 1000\,d$). Periodic boundary conditions are used in the $x$ (wind) direction. The results presented here are obtained for a density ratio $\rho_p/\rho_f=500$, a grain Reynolds number $\mathcal{R}=d/\nu \sqrt{(\rho_p/\rho_f -1) gd}=22$ ($\nu$ is the air kinematic viscosity) and shear velocities in the range $u_*/u_{\rm th} = 1$--$5$. The colored background codes for the wind velocity, see wind profile (left). (b) Close-up view at the scale of the ripple wavelength, featuring saltation trajectories, with hop-height between $15$ and $30d$ . The average resonant trajectory is shown in red. (c) Zoom at the level of the interfacial. A collision between a grain in saltation (orange) and a grain in reptation (green) is sketched.}
	\label{fig1}
\end{figure*}

\section{Results}

\subsection{Sand ripple instability}
\begin{figure*}
\centerline{\includegraphics{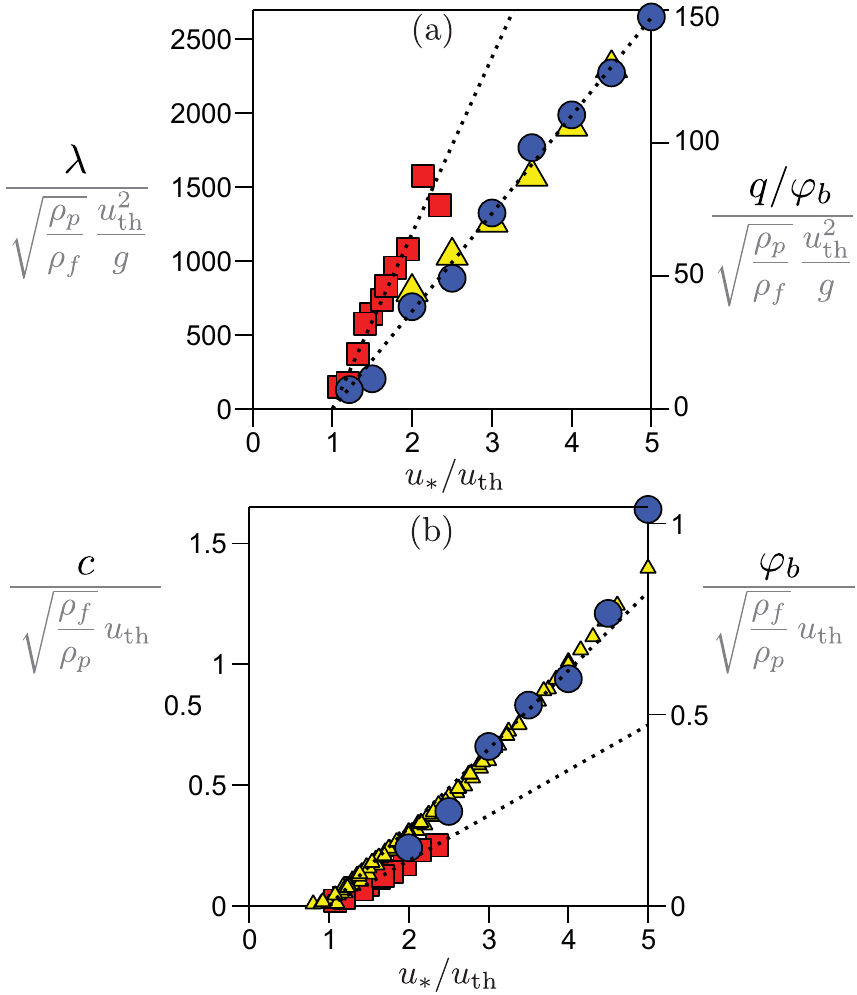}}
\caption{Selection of ripple wavelength and propagation speed. Blue circles: data consistently deduced from the most unstable mode of the dispersion relation, and measured starting from a flat sand bed. Red squares: experimental data (8). (a) Ripple wavelength $\lambda$ as a function of the wind shear velocity $u_*$ (left axis). Yellow triangle: ratio of the saltation flux $q_s$ to the erosion/deposition rate $\varphi_b$ (right axis). (b) Propagation velocity $c$ as a function of $u_*$ (left axis). Yellow triangles: erosion/deposition rate $\varphi_b$, measured for many values of density ratios $\rho_p/\rho_f$ and grain Reynolds numbers Re (right axis). Dashed lines: linear fits to the data. Statistical error bars are of the size of the symbols. }
	\label{fig2}
\end{figure*}

Starting from a flat sediment bed, disturbed only by the randomness in the granular arrangement, one observes in the simulations the emergence and the propagation of ripples (Fig.~\ref{fig1}, Fig.~S1 and Video~S2). {Tracking the grain trajectories, one can see that the saltation rain above the rippled bed is strongly modulated (Fig.~1b). As observed in experiments (Fig.~S8), grains in saltation preferentially hit the bed upwind of the ripple crests. As ejected grains make small hops, the reptation flux tends to be enhanced on the windward side and reduced on the lee side. This results into a net transport from the troughs towards the crests that amplifies topographic disturbances, hence the instability.} The spontaneous ripple pattern has a wavelength $\lambda$ and a propagation speed $c$, both varying linearly with the imposed wind shear velocity (Fig.~\ref{fig2}), in agreement with experimental observations \cite{S63,SL78,ACP06}. Both $\lambda$ and $c$ are found to vanish when $u_*$ tends to the threshold value $u_{\rm th}$, above which sediment transport takes place. The key issue addressed in this article is the origin of these scaling laws, which results from the interplay between saltation and reptation transport modes.

To investigate quantitatively the dynamical mechanisms leading to the ripple instability, we have also performed simulations starting from a modulated bed whose topography follows a sinusoidal profile of given wavenumber $k$ and of small initial amplitude $|\hat Z|(0)$. The phase and the modulation amplitude $|\hat Z|(t)$ are measured as a function of time, by a simple Fourier transform of the elevation profile at the wavenumber $k$. As expected for a linear instability, the growth or the decay of the disturbance can be fitted to an exponential of the form $|\hat Z|(t) = |\hat Z|(0) e^{\sigma t}$ (Fig. S1) which gives the growth rate $\sigma$. The resulting dispersion relation $\sigma(k)$, obtained for each wind velocity, is typical of a long-wave instability (Fig.~\ref{fig3}a): small wavenumbers (large wavelengths) are unstable ($\sigma > 0$) while large wavenumbers (small wavelengths) are stable ($\sigma < 0$). The most unstable mode, determined by the wavenumber that maximizes $\sigma$, coincides with the wavenumber $2 \pi/\lambda$ of the pattern that spontaneously emerges from a flat sediment bed.

\subsection{Destabilizing effect of reptation}
We have determined the contribution of the grains in reptation to the growth rate by selecting the grains with a hop height smaller than $3d$ and measuring the difference between deposition and erosion rates.  As shown in Fig.~\ref{fig3}b, we find that reptation has a destabilizing effect ($\sigma > 0$) and contribute to $\sigma(k)$ linearly in $k$. {The ratio $\sigma/k$ is homogeneous to an erosion/deposition rate and has therefore the dimension of a velocity. This linear scaling in $k$ must then originate from a characteristic value of such a rate associated with reptation. To determine it, we have measured the vertical flux density profile $\varphi(z)$, defined as the volume of the grains crossing a unit horizontal surface at altitude $z$ per unit time ({\it SI text}). $\varphi(z)$ systematically presents a maximum at the surface of the static bed (Fig.~S3), which defines the basal erosion/deposition rate $\varphi_b$. The scaling properties of sand ripples directly originate from an unexpected dependence of $\varphi_b$ on the wind speed, which must therefore be discussed in details.}

{Importantly, the vertical flux density profile $\varphi(z)$ reveals the existence of a yet unnoticed interfacial layer separating the saltation zone from the static bed (Fig.~S3). In this layer, which is a few grain sizes thick, the grain volume fraction is close to that of the static bed and mid-air collisions are frequent \cite{SMcE96,PJ05,DHL05,CAPH13}. As evidenced by the large collision probability in this layer (Fig.~S3a), only a small fraction of the grains arriving from the upper transport layer truly impacts the static bed: most of them actually bounce back before. Because of these collisions, the shear stress carried by the grains is transferred from a kinetic form, i.e. a flux of momentum associated with a particle flux, to a contact stress. As the typical grain velocity in this layer is set by $\varphi_b$, the associated collisional stress scales as $\sim \rho_p \varphi_b^2$. In a situation of steady and homogeneous transport, this stress must balance the grain-borne shear stress in the upper transport layer, which scales with $\rho_f$ and with the square of the excess shear velocity $\delta u = u_* - u_{\it th}$. The basal erosion/deposition rate therefore varies as
\begin{equation}
 	\varphi_b \sim \sqrt{\rho_f/\rho_p} (u_* - u_{\it th}),
\end{equation}
in agreement with the numerical data computed with different values of the ratios $u_*/u_{\rm th}$, and $\rho_p/\rho_f$ ({\it SI text}) (Fig.~\ref{fig2}b). This scaling law is found to hold even close to the threshold (Fig.~S3b), which means that the ejection process and the redistribution of momentum in the interfacial layer must involve a large number of grains, even when splashing grains are well separated in time. This observation challenges the common view on aeolian sediment transport and calls for the seek of collective effects in the splash process.}

{The grain hop length distribution $\psi(\ell)$ ({\it SI text}) reflects the dynamical mechanisms dominating each of these transport layers. While $\psi$ decreases with $\ell$ as a stretched exponential in the upper layer, it behaves as a scale-free power-law $\psi_b(\ell)\sim\varphi_b \ell^{-1}$ in the interfacial layer (Figs.~S4,S5). This, surprisingly, means that there is no characteristic hop length (a given number of grain sizes) for grains departing from the static bed -- for reptation in particular -- while there is one for grains flying in the upper transport layer (several tens of $d$). Grains ejected from the static layer inherit their scaling laws from the impacting grains. The scale-free behavior observed in the interfacial layer is another indication of the collective processes at work. As an important consequence, the erosion/deposition rate $\varphi_b$ is the single characteristic quantity of this layer. Accordingly, we have found that, varying the wind velocity and the density ratio $\rho_p/\rho_f$ in the simulations, the contribution of reptation to the growth rate $\sigma$ always takes the form $a \varphi_b k$, with multiplicative factor $a$ of order 1 (Fig.~\ref{fig3}b).}

\subsection{Stabilizing effect of saltation}
\begin{figure*}
\centerline{\includegraphics{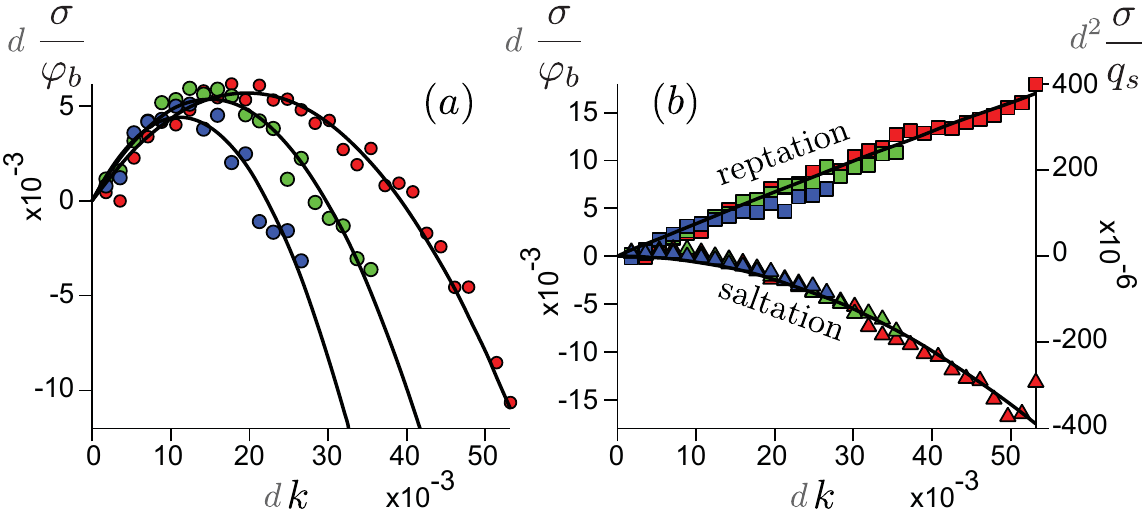}}
\caption{Dispersion relation measured from the numerical simulations. (a) Ripple growth rate $\sigma$ as a function of the wavenumber $k=2\pi/\lambda$ for different wind speeds (red: $u_*/u_{\it th} = 3$, green: $u_*/u_{\it th} = 4$, blue: $u_*/u_{\it th} = 5$). The solid line is the fit of Eq. (3). (b) Contributions of grains in saltation (right axis) and grains in reptation (left axis) to $\sigma$. Same color code as in panel (a). }
	\label{fig3}	
\end{figure*}

{The contribution of grains in saltation (i.e. grains with a hop height larger than $3d$) to the growth rate $\sigma$ is found to be negative and quadratic in $k$ (Fig.~\ref{fig3}b). Varying $u_*/u_{\rm th}$ and $\rho_p/\rho_f$, we can further establish that the saltation-induced growth rate has the form $-b q_s k^2$, where $q_s$ is the total sediment flux over a flat bed and the multiplicative factor $b$ is of order 1 (Fig.~\ref{fig3}b). Saltation has therefore a direct stabilizing effect consistent with a topographic diffusion, i.e. a downslope component of the saltation flux. As evidenced in experiments \cite{IR99}, the overall sand transport -- not only reptation -- is sensitive to the bed slope, with a diffusion coefficient  proportional to the total sediment flux $q_s$. This slope effect can be understood from a momentum balance. Most of the sediment flux takes place in the upper part of the transport layer and is controlled by the transfer of momentum from the wind to the grains. At steady state, this momentum is balanced by resistive forces due to collisions with the bed. Since these forces are effectively smaller downslope than upslope, the surplus of momentum from the wind can set more grains into motion and thus increases the flux downslope.}

The measurement of $q_s$ in the simulations ({\it SI text}) shows a quadratic dependence on $u_*$ (Fig.~S2), in agreement with controlled experiments \cite{IR99}:
\begin{equation}
q_s \sim \frac{\rho_f u_{\it th}}{\rho_p g} (u_* - u_{\it th})^2.
\end{equation}
This scaling behavior is fairly well established and has been interpreted as a consequence of the feedback of the grains on the wind flow. This feedback keeps the characteristic saltation velocity constant and proportional to the shear velocity threshold $u_{\it th}$ \cite{B41,Cetal09,DCA11,KPMK12,DAC12}, a necessary condition to reach steady state sediment transport when one impacting grain is replaced, on average, by a single ejected grain \cite{UH87}.

\subsection{Dispersion relation}
Summing up the contributions of reptation and saltation, the dispersion relation $\sigma(k)$ is well fitted, for all shear velocities $u_*$, by the parabolic function
\begin{equation}
	\sigma = a \varphi_b k - b q_s k^2,
\end{equation}
with multiplicative constants, $a$ and $b$ of order 1. This constitutes a major difference with existing models, which predict that both the destabilizing and the stabilizing terms grow like $k^2$ at small wavenumber. The most unstable wavelength computed from the dispersion relation is proportional to the sediment transport length, defined as the ratio of the horizontal flux to the erosion/deposition rate $q_s/\varphi_b$ (Fig.~\ref{fig2}a). Because $q_s$ is quadratic while $\varphi_b$ is linear with respect to the shear velocity, we get:
\begin{equation}
	\lambda \sim \frac{q_s}{\varphi_b} \sim \sqrt{\frac{\rho_f}{\rho_p}}\frac{u_{\it th}}{g}(u_* - u_{\it th}) .
\end{equation}
We have checked that this scaling law is robust to the value of the grain to fluid density ratio $\rho_p/\rho_f$ and of the grain Reynolds number. Although our system is two-dimensional and composed of rather soft particles, the emergent wavelength in the simulations is only a factor 1.8 away from that obtained in wind tunnel experiments (Fig.~\ref{fig2}a).

In the field, initial ripples develop towards a statistically steady pattern. Their wavelength eventually results from fluctuating wind conditions and from ripple non-linear interactions. However, measurements show that the linear dependence of the wavelength on the wind velocity still holds for developed ripples \cite{ACP06}. Furthermore, Martian ripple wavelength (in the range $2$--$6$~m, see Table~S1 and Fig.~S6) is $\simeq 30$ times larger than on a typical earth sand dune. Under the assumption that the most probable wind velocity for ripple formation is the static transport threshold, our simulations effectively predict that for these conditions the ratio $q_s/\varphi_b$ is ~20 times larger on Mars than on Earth, due to different values of atmospheric characteristics (density, viscosity) and gravity. 

Experimental data and numerical simulations also agree on the scaling law obeyed by the propagation speed of ripples (Fig.~\ref{fig2}b):
\begin{equation}
	c \sim \varphi_b \sim \sqrt{\rho_f/\rho_p}(u_*-u_{\it th}).
\end{equation}
The product of the wavelength and the speed is therefore proportional to the total sediment flux $q_s \sim c \lambda$. This relation has been successfully tested in the field, where the three quantities $\lambda$, $c$ and $q_s$ have been measured independently (Fig.~S7). This opens the promising perspective to track sand mass transfers in real-time, by performing remote measurements of sediment fluxes from aerial views of sandy area, on Mars in particular (Fig.~S6).

\begin{figure*}
\centerline{\includegraphics{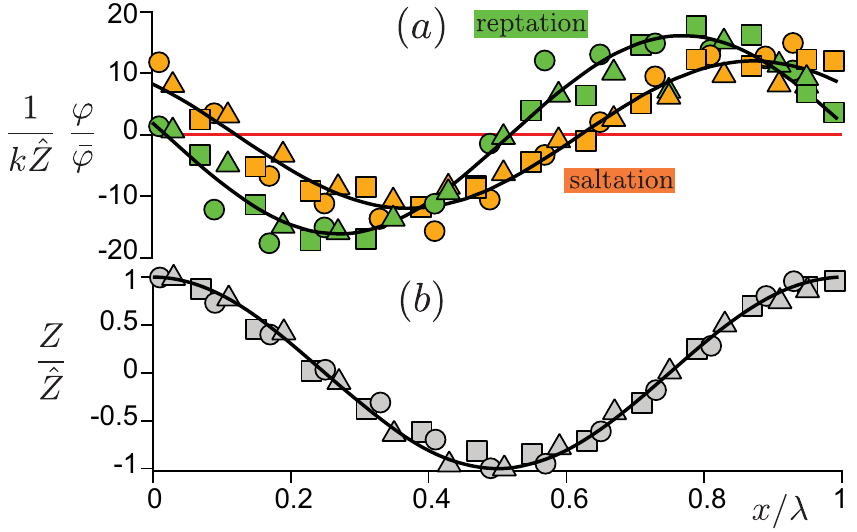}}
\caption{Focusing of saltating trajectories. (a) Rescaled vertical flux density profiles of grains in reptation (green symbols) and grains in saltation (orange symbols) for three different wind velocities (squares: $u_*/u_{\it th} = 3$, circles: $u_*/u_{\it th} = 4$, triangles: $u_*/u_{\it th} = 5$). Fluxes are evaluated on horizontal lines to subtract the screening effect ({\it SI text}). The solid red line corresponds to a uniform saltation rain. (b) Normalized ripple elevation profiles. Black solid lines: sinusoidal fits.}
	\label{fig4}	
\end{figure*}


\section{Discussion}
The origin of the ripple instability can be understood as follows. As previously described, grains are eroded from the troughs and accumulate on the crests because of the modulation of the reptation flux, which is maximal on the windward side and minimal on the lee side (Fig.~\ref{fig4}). In contrast with the geometric screening scenario, simulations and experiments show that the saltation rain is also strongly modulated (Fig.~\ref{fig4} and Fig.~S8). The impacting flux is nonetheless larger on upwind-slopes due to the solid angle effect, but also because of a stochastic focusing of the trajectories. Indeed, the rate at which grains leave the bed is modulated similarly to the impacting rate: if there are more grains arriving at the upwind-slope inflexion point, there are also more grains departing from that position. The key point is that these grains must statistically be the same: the modulation of sediment transport is essentially due to grains departing from the inflexion point and arriving at the following inflexion point located downwind. The consequence for the ripple emergence is fundamental: the modulation of the reptation flux is driven by resonant trajectories, those whose hop length is comparable to the wavelength $\lambda$. Since processes at work in the interfacial layer are scale-free, the reptating grains that contribute the most to the instability have hop lengths proportional to $\lambda$, and not to $d$: they inherit their characteristic scale from that of saltating grains. The new picture proposed here therefore substantially changes the paradigm of aeolian transport.

\vspace*{0.3cm}
\noindent
\rule[0.1cm]{3cm}{1pt}

This study was supported by an ANR grant Zephyr. We thank H. Elbelrhiti and O. Pouliquen for discussions and assistance with the field work. We are grateful to A.B. Murray, J. Tavacoli and G. Wiggs for a careful reading of the manuscript and useful comments.


\newpage
\appendix

\section{Supplementary method}

The numerical model is based on \cite{DAC12}, where discrete element method (DEM) for particles are coupled to a continuum description of hydrodynamics based on Reynolds averaged Navier-Stokes equations (RANS) closed at the first order. The system is a (2+1)D spatial domain and periodic boundary conditions are used in the $x$ (flow) direction. Hydrodynamics is solved both in the dilute transport layer and in the granular bed. The fluid, characterized by a density $\rho_f$ and a viscosity $\nu$, is set into motion by a turbulent shear stress $\rho_f u_*^2$ imposed far from the bed. The grains, of mean diameter $d$ and density $\rho_p$ are subject to gravity $g$ and interact through contact forces. The two-way coupling occurs by means of drag and Archimedes forces, incorporated at the discrete level in the equations of motion of the grains, and at the continuum level via a body force term in the RANS equations.

\subsection{Control parameters}

The control parameters of the model are the density ratio $\rho_p/\rho_f$, the grain Reynolds number $\mathcal{R}=d/\nu \sqrt{\left(\rho_p/\rho_f -1 \right) gd}$, and the shear velocity $u_*$. The dynamical transport threshold below which the saturated sediment flux vanishes is characterized by $u_{\rm th}$. The results presented here are obtained for $\rho_p/\rho_f=500$, $\mathcal{R}=22$ and $u_*/u_{\rm th} = 1$--$5$. To check the dependence of the results on the density ratio and on the grain Reynolds number, additional runs for $\mathcal{R}=1,11$ and for $\rho_p/\rho_f=100, 250, 500, 750, 1000$ have been performed. Microscopic grain parameters (stiffness, restitution coefficient, contact friction) are chosen in a range where they do not affect the results.

\subsection{Forces on particles}

We model the contact forces following a standard approach in DEM codes (see e.g. \cite{R11} and references therein), where normal and tangential components are described by spring dash-pot elements. A microscopic friction coefficient is also introduced. We assume that the net hydrodynamical force acting on a particle due to the presence of the fluid is dominated by the sum of the drag and Archimedes forces:
\begin{eqnarray}
\vec f^p_{\rm drag} & = & \frac{\pi}{8} \rho_f d^2 C_d(R_u) |\vec u - \vec u^p| (\vec u - \vec u^p), \label{fdrag}\\
\vec f^p_{\rm Arch} & = & \frac{\pi}{6}d^3 {\rm div} \sigma^f. \label{fArch}
\end{eqnarray}
$\vec u^p(x,z)$ is the grain velocity and $\vec u(z)$ is the fluid velocity at grain's height $z$.  $R_u = |\vec u - \vec u^p| d/\nu$ is the particle Reynolds number $R_u$ based on the velocity difference between the particle and the fluid. $\nu$ is the fluid viscosity. For the drag coefficient $C_d$, we use the phenomenological expression \cite{FC04}: $C_d(R_u) = \left(\sqrt{C^{\infty}_d} + \sqrt{R_u^c/R_u} \right)^2$, where $C^{\infty}_d \simeq 0.5$, is the drag coefficient of the grain in the turbulent limit ($R_u \rightarrow \infty$), and $R_u^c \simeq 24$ is the transitional particle Reynolds number above which the drag coefficient becomes almost constant. $\frac{\pi}{6} d^3$ is the grain volume and $\sigma^f_{ij} = -p^f\delta_{ij} + \tau^f_{ij}$ is the undisturbed fluid stress tensor (written in terms of the pressure $p^f$ and the shear stress tensor $\tau^f_{ij}$). In first approximation, the stress is evaluated at the center of the grain. The lift force, lubrication forces and the corrections to the drag force (Basset, added-mass, Magnus, etc.) are neglected.

\subsection{Hydrodynamics and coupling}
Hydrodynamics is described by the Reynolds averaged Navier-Stokes equations, which in the steady and homogeneous case simplify into
\begin{eqnarray}
	\dr_z p^f & = & - \rho_f g, \label{Eqpf}\\
	\dr_z \tau^f_{xz} & = & \dr_z \tau^p. \label{Eqtauf}
\end{eqnarray}
Eq.~\ref{Eqtauf} integrates as $\tau^f_{xz}(z) = \rho_f u_*^2- \tau^p(z)$, where the shear velocity $u_*$ is defined by the undisturbed (grain free) wall shear stress and the grain borne shear stress $\tau^p(z)$ is defined by the average sum of the hydrodynamic force acting on all grains moving above $z$ over an area $A$ (given by the total horizontal extent of the domain):
\begin{equation}
\tau^p(z) = \frac{1}{A} \left < \, \sum_{p\in \{z^p > z \}} \left( f_{{\rm drag}, x}^p + f_{{\rm Arch}, x}^p \right) \right >,
\label{defFx}
\end{equation}
where the symbols $\langle . \rangle$ denote ensemble averaging.

The fluid borne shear stress is related to the average fluid velocity field by a Prandtl-like turbulent closure with a mixing length $L$:
\begin{equation}
\label{eq:tauf_u}
\tau^f_{xz} = \rho_f (\nu + L^2 |\dr_{z} u_x|)\dr_{z} u_x.
\end{equation}

When sand ripples develop, the bed becomes gently undulated. In first approximation, for small bed slopes, the equilibrium equations \ref{Eqpf} and \ref{Eqtauf} are still valid, once expressed in the relative coordinates $(x,z-Z(x,t))$, where the bed elevation profile $Z(x,t)$ is calculated around a given $x$ from mass conservation:
\begin{equation}
\label{eq:BedSurface}
Z(x) = \frac{1}{\phi_b} \int_0^{\infty} \left[ \phi(x,z) -  \frac{1}{A} \int_0^A \phi(x,z) \, \dd x \right ] \dd z,
\end{equation}
here $\phi(x,z)$ is the grain volume fraction within a box $(\dd x,\dd z)$ around $(x,z)$:
\begin{equation}
\phi(x,z) = \frac{1}{\dd x \dd z} \sum_{p\in \{x;x+\dd x\}\times\{z;z+\dd z\}} \frac{\pi}{6} d^3,
\end{equation}
and $\phi_b$ is the average volume fraction inside the bed. Note that the average of $Z$ over $x$ is zero by construction.

\subsection{Sediment flux and erosion/deposition rate}
The equilibrium sediment flux $q_s$ is calculated in the simulations as:
\begin{equation}
q_s = \frac{1}{\phi_b A} \, \frac{\pi}{6}d^3 \sum_p u^p,
\label{computeqs}
\end{equation}
where the sum is over all particles. The vertical flux density $\varphi(z)$ at altitude $z$ is calculated in the simulations as:
\begin{equation}
\varphi = \frac{1}{\phi_b A\dd z} \, \frac{\pi}{6}d^3 \sum_{p \in (z,z+\dd z)} u^p_\uparrow,
\label{computevarphib}
\end{equation}
where $u^p_\uparrow$ is the grain upward velocity and the sum is over grains within a layer of thickness $\dd z$ above $z$. The erosion/deposition rate $\varphi_b$ is defined as the maximum value of $\varphi$, which is attained at the static bed (z=0).

For the data displayed in Fig. 4, corresponding to the variations of the vertical flux density along the ripple profile, $\varphi$ must be computed for each position $x$. In order to remove the solid angle effect due to the bed topography, the control volume over which the sum of Eq. \ref{computevarphib} is computed is horizontal and \emph{not} inclined parallel to the bed surface.

\subsection{Hop length distributions}
Similarly to the vertical flux density $\varphi(z)$, the distribution of hop length $\psi(\ell)$ is determined at different altitudes $z$. It is measured by counting the volume of grains  that cross  a unit surface per unit time  after a hop of length $\ell$. By definition, the integral of $\psi(\ell)$ at altitude $z$ is $\varphi(z)$. We have measured the distribution $\psi_s(\ell)$ in the upper transport layer at a reference altitude $z=5d$, i.e. located above the interfacial layer. We have also measured the distribution $\psi_b(\ell)$ in the interfacial layer, i.e. measured at the surface of the bed.


\section{List of symbols in the main paper}

\noindent
\begin{tabular}{ll}
gravity acceleration 								& $g$ \\
atmosphere density								& $\rho_f$ \\
grain density									& $\rho_p$ \\
grain diameter									& $d$ \\
& \\
shear velocity									& $u_*$ \\
associated shear stress							& $\rho_f u_*^2$ \\
shear velocity at impact threshold					& $u_{\rm th}$ \\
associated threshold shear stress					& $\rho_f u_{\rm th}^2$ \\
excess shear velocity							& $\delta u = u_* - u_{\rm th}$ \\
contact stress in the interfacial layer					& $\rho_p u_b^2$ \\
& \\
saltation flux									& $q_s$ \\
vertical flux density								& $\varphi$ \\
basal erosion/deposition rate						& $\varphi_b$ \\
& \\
ripple wavelength								& $\lambda$ \\
ripple growth rate								& $\sigma$ \\
ripple propagation velocity						& $c$ \\
& \\
grain hop length								& $\ell$ \\
hop length distribution							& $\psi(\ell)$ \\
modulation rate of the impacting flux					& $\mathcal{A}(\ell)$
\end{tabular}

\newpage

\section{Supplementary table and figures}

\begin{figure*}
\centerline{\includegraphics{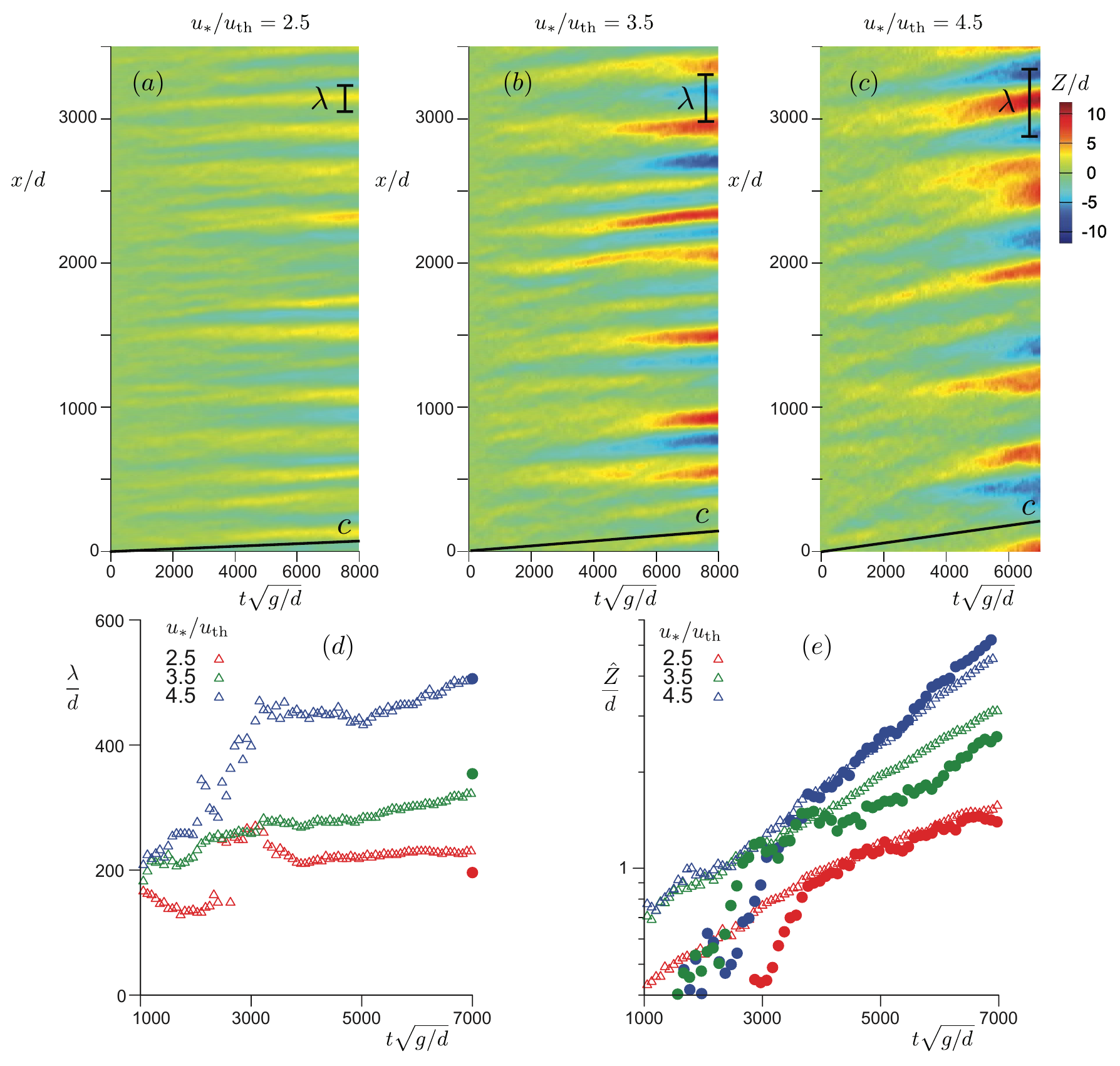}}
	{{\bf Fig. S1}. (a-c) Space-time diagram showing the formation of ripples, starting from a flat sediment bed (numerical data). Color code (see legend): variation of the bed elevation. Average wavelengths  $\lambda$ (scale bar) and speeds $c$  (solid line slope) are indicated for reference. (d-e) Time evolution of the bed profile wavelength $\lambda$ and amplitude $\hat Z$, either calculated from the autocorrelation function (open symbols) or from the fastest growing mode of the Fourier decomposition (filled symbols). Data corresponding to three different wind shear velocities are displayed, see legends in all panels.}
	\label{S1}
\end{figure*}

\begin{figure*}
\centerline{\includegraphics{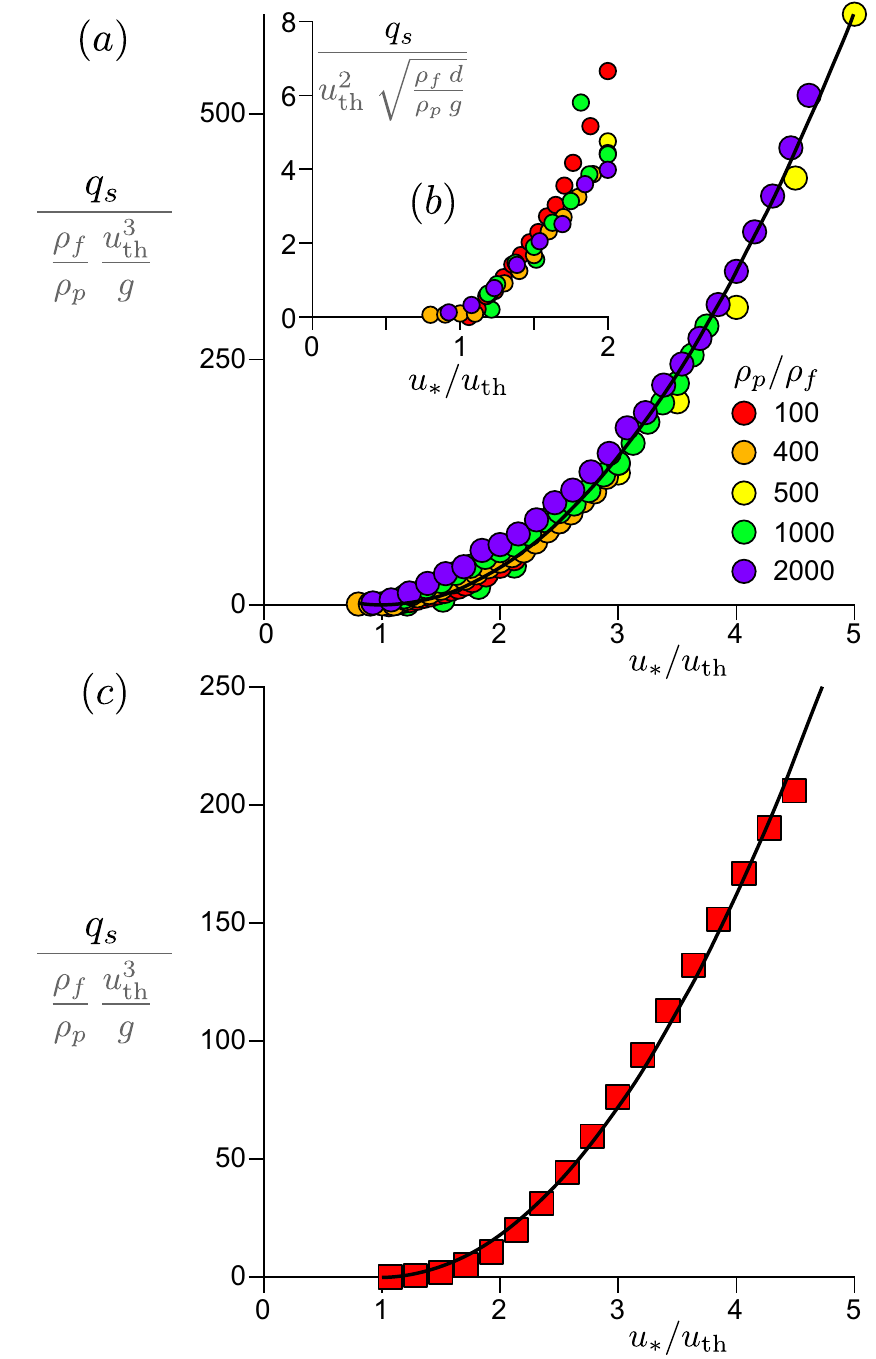}}
	{{\bf Fig. S2}. Sediment flux $q_s$ as a function of the shear velocity $u_*$ rescaled by the threshold shear velocity $u_{\rm th}$. (a-b) Numerical data, rescaled to show the dependence with respect to $\rho_p/\rho_f$ (see legend) close to the threshold (b) and far from it (a). (c) Experimental data obtained in a wind tunnel, from Iversen \& Rasmussen (1999). Reproducibility is achieved by these authors within $5\%$, which corresponds roughly to the size of the symbols. Solid lines: fit by Eq.~1 from main text.}
	\label{S2}
\end{figure*}

\begin{figure*}
\centerline{\includegraphics{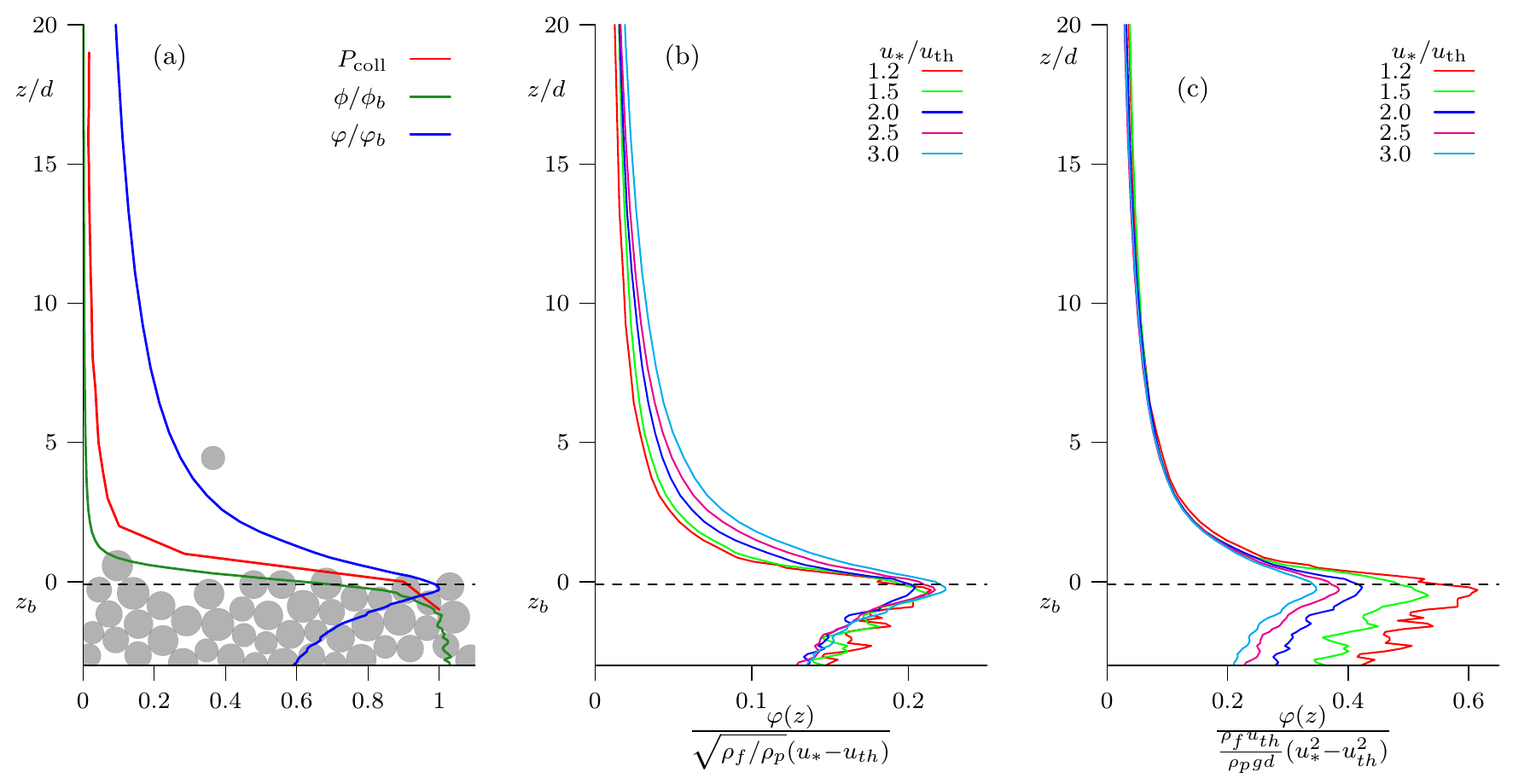}}
	{{\bf Fig. S3}. (a) Typical vertical profiles of three quantities: the volume fraction $\phi$ rescaled by its value at the bed $\phi_b$, the vertical flux density $\varphi$, rescaled by the erosion/deposition rate $\varphi_b$, and the collision probability density $P_{\rm coll}$ . (b) Profile of the vertical flux density $\varphi(z)$ for different shear velocities (see legend), rescaled to show the linear dependence on $u_*$ in the interfacial layer. (c) Profile of the vertical flux density $\varphi(z)$, rescaled to show the quadratic dependence on  $u_*$ in the upper transport layer. The reference altitude $z=0$ is chosen at the maximum of $\varphi(z)$, which defines the erosion/deposition rate $\varphi_b=\varphi(0)$.}
	\label{S3}
\end{figure*}

\begin{figure*}
\centerline{\includegraphics{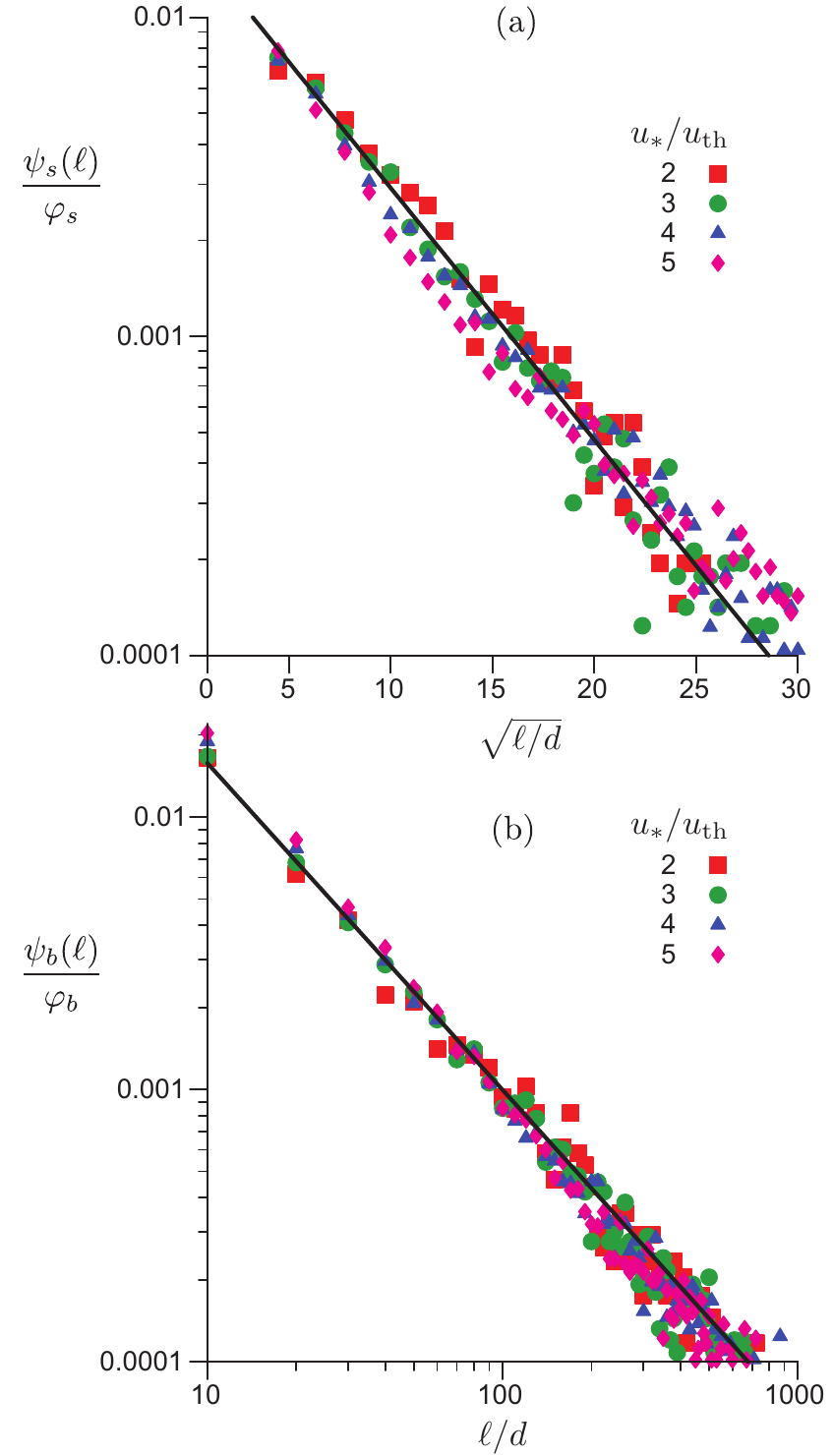}}
	{{\bf Fig. S4}. Hop length distributions (a) in the upper transport layer i.e. measured at $z=5d$ and (b) in the interfacial layer i.e. measured at the surface of the bed. The different symbols correspond to different wind speeds (see legends). Solid line in (a): $\psi_s(\ell)=\varphi_s \ell_s^{-1} \exp \left(-\sqrt{\ell/\ell_s}\right)$, with $\ell_s\simeq 30 d$ independent of the wind speed. Solid line in (b):  $\psi_b(\ell)=\varphi_b/d \left(d/\ell\right)^\alpha$, with $\alpha\simeq 1.2$ over the range  $10d$-$1000d$. In the main text, we use the convenient approximation  $\alpha=1$.}
	\label{S4}
\end{figure*}

\begin{figure*}
\centerline{\includegraphics{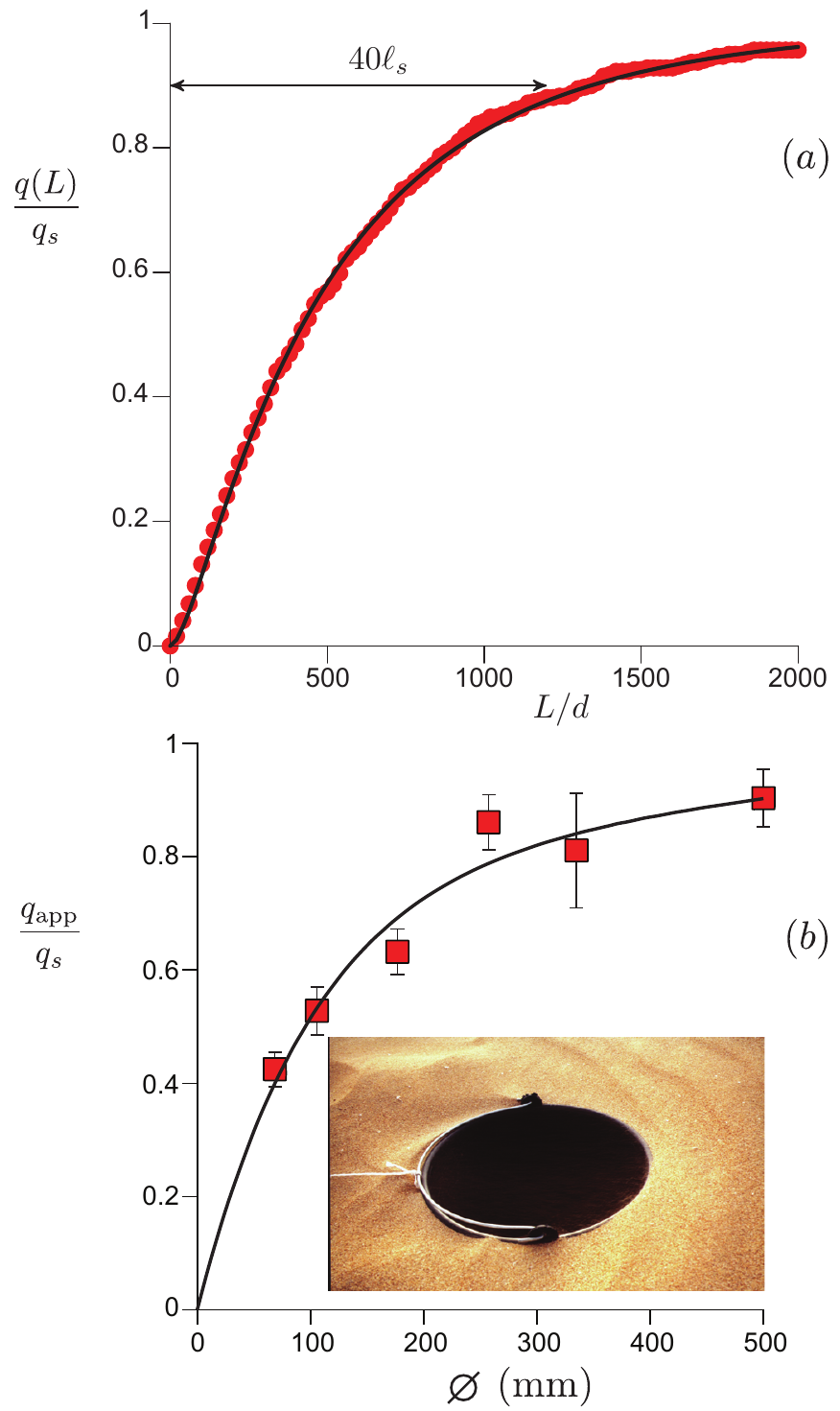}}
	{{\bf Fig. S5}. (a) Contribution to the sand flux of trajectories whose hop length verifies $\ell<L$. Numerical simulations (symbols) and analytical approximation (solid line) using $\psi_s(\ell)$, see Fig.~S4. (b) Sand flux  $q_{\rm app}$ measured in a circular trap (see inserted photo) as a function of the trap diameter (field data). $q_{\rm app}$ is normalised by the saturated flux $q_s$. The solid line is the theoretical expectation, after integration of the distribution $\psi_s(\ell)$. The single fitting parameter is $\ell_s \simeq 13\;{\rm mm} \simeq 70 d$. The statistical error bars are estimated by performing the same measurements $10$ times.}
	\label{S5}
\end{figure*}

\begin{figure*}
\centerline{\includegraphics{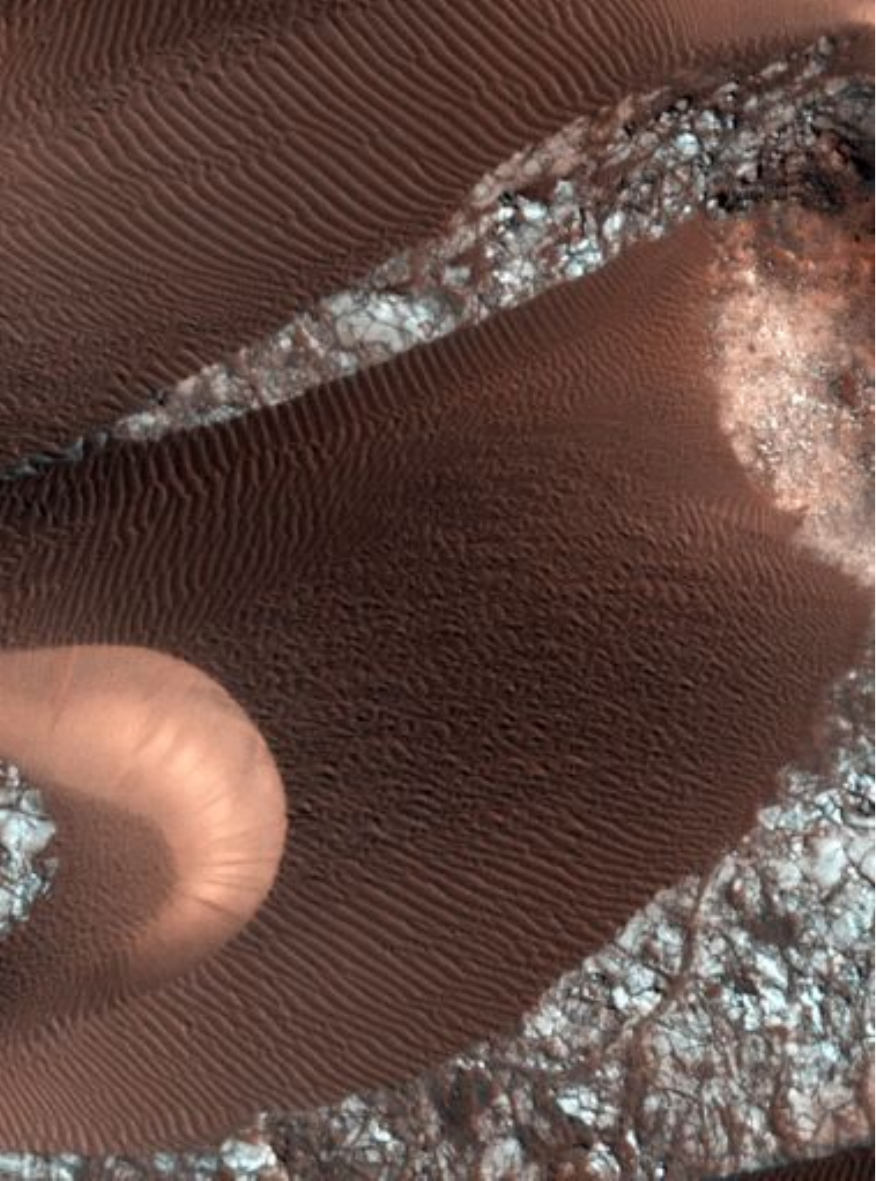}}
	{{\bf Fig. S6}. Aerial photography ($280$ m wide) showing sand ripples over a martian dune (see the avalanche slip-face on the left) in Nili Patera (Image credit: NASA/JPL/University of Arizona).}
	\label{S6}
\end{figure*}

\begin{figure*}
\centerline{\includegraphics{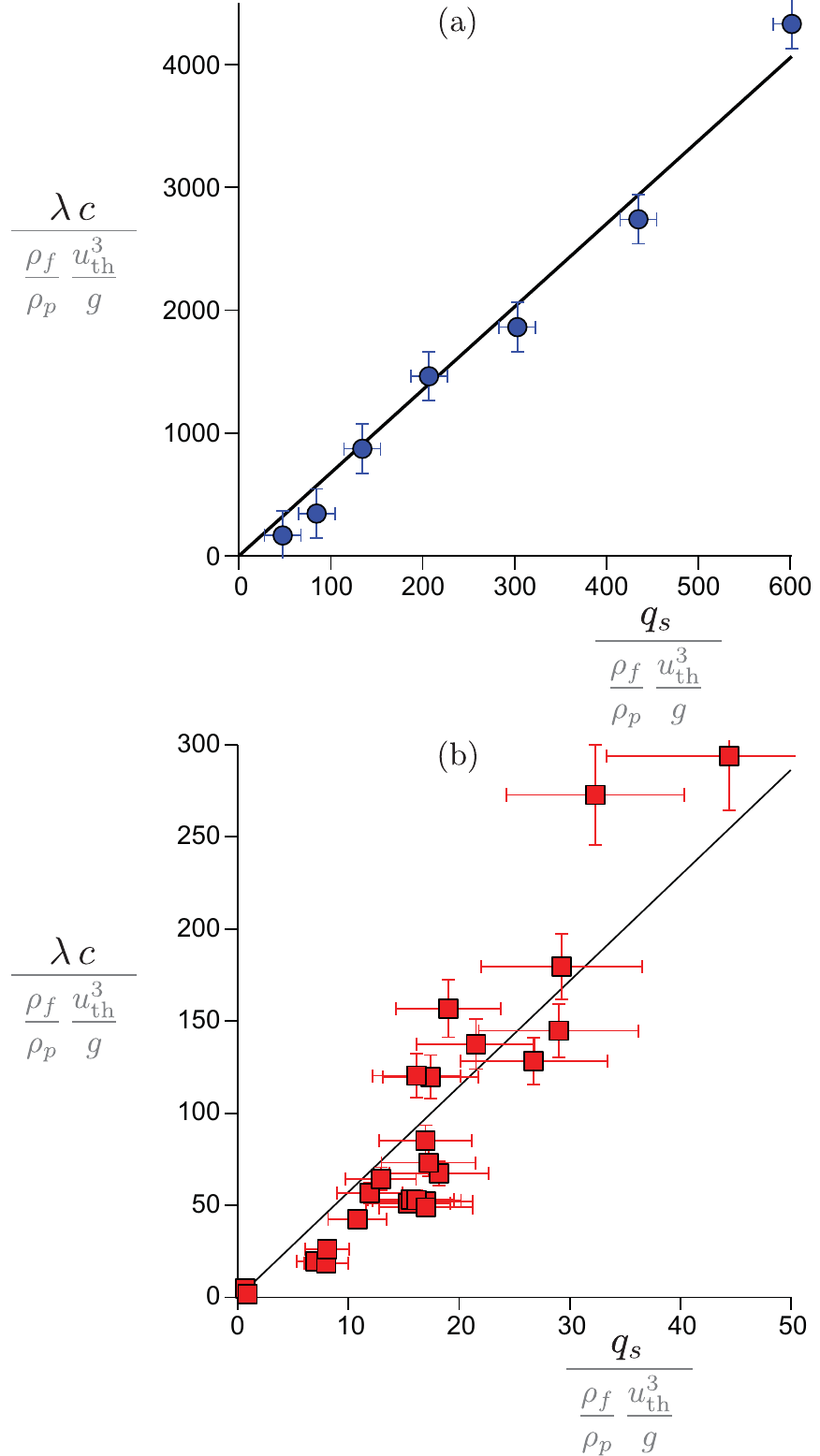}}
	{{\bf Fig. S7}. Product of the ripple wavelength $\lambda$ and the ripple velocity $c$ plotted against the grain flux $q_s$. Wind tunnel (a) and field (b) data, from Andreotti et al. (2006). In both cases, the statistical error bars, deduced from the dispersion of independent measurements, are displayed. The large dispersion in the field measurements may result from systematic effects, as the wavelength is not always adapted to the wind, whose characteristic strength fluctuates.}
	\label{S7}
\end{figure*}

\begin{figure*}
\centerline{\includegraphics{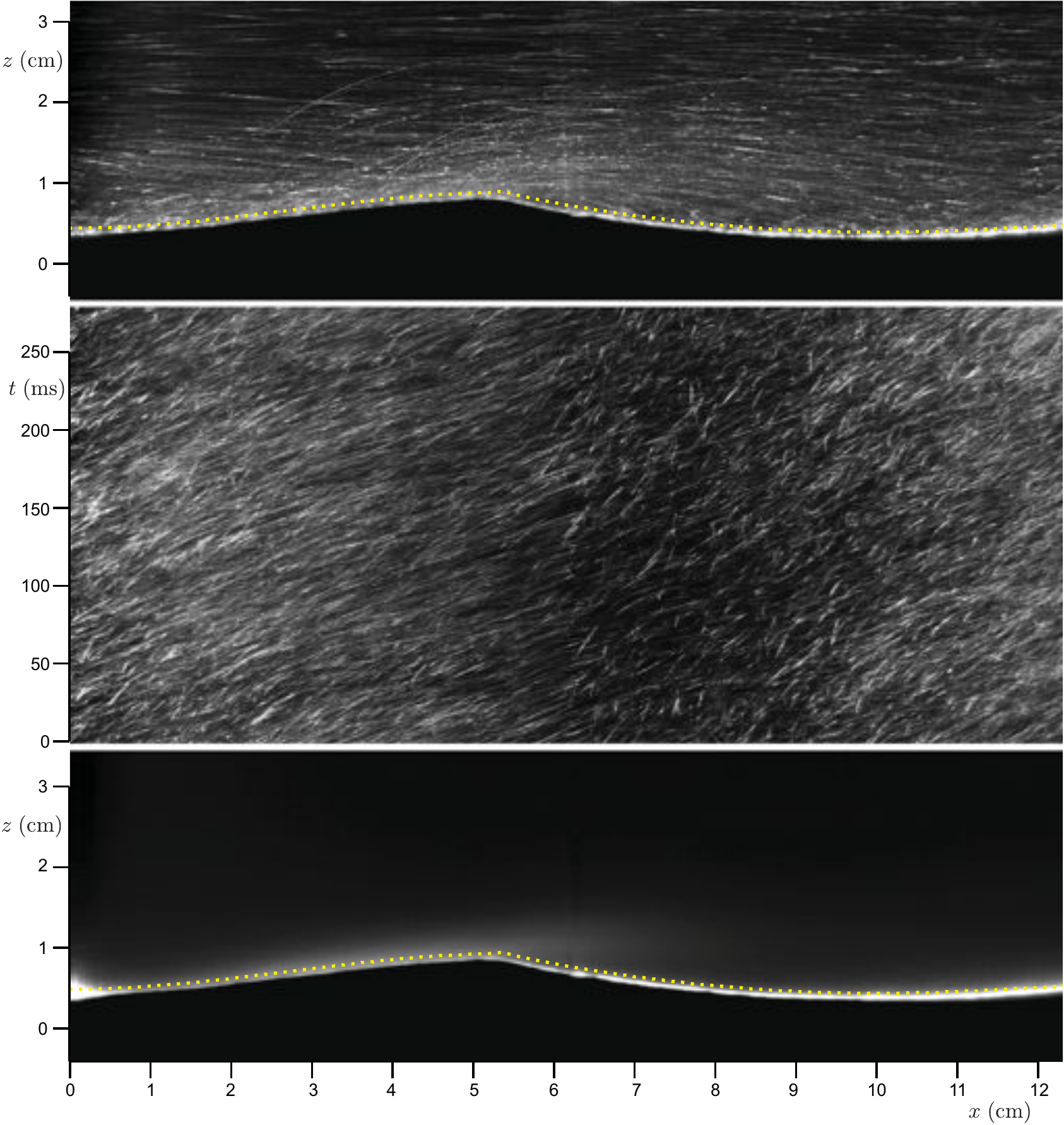}}
	{{\bf Fig. S8}. Experimental evidence of a modulated saltation flux. Images obtained after analysis of a fast video movie recorded in a wind tunnel. The ripple is vertically illuminated with a laser sheet. Images are corrected from the heterogeneity of the light intensity. Top panel: image obtained by determining the maximum intensity at each pixel, over a certain time window. Grain trajectories can be observed. Middle panel: space-time diagram showing the gray level measured along the yellow dotted line and plotted as a function of time. The slope of the streaks give the velocity. One directly visualizes the modulation of the impacting flux. Bottom panel: average image, showing the map of the grain volume fraction. The deficit of saltating grains in the trough is directly visible.}
	\label{S8}
\end{figure*}

\begin{table}[h]

\begin{tabular*}{\hsize}{@{\extracolsep{\fill}}lll}
\hline
Parameter										& Earth				& Mars							\cr 
\hline 
Gravity acceleration	$g$ (m$^2$/s)						& $9.8$ 				& $3.7$							\cr 
Atmosphere density $\rho_f$ (kg/m$^3$)					& $1.2$				& $1.5$--$2.2 \, 10^{-2}$				\cr 
Atmosphere viscosity $\nu$  (m/s$^2$)					& $1.5 \, 10^{-5}$		& $6.3 \, 10^{-4}$					\cr 
Grain diameter $d$ ($\mu$m)							& $165$--$185$ (a)		& $100$--$200$					\cr 
Grain density $\rho_p$ (kg/m$^3$)						& $2650$				& $3000$							\cr 
Static to dynamic threshold velocity ratio $u_*/u_{\rm th}$		& $1.8$ (b) 			& $3$--$9$ (c) 						\cr 
Grain Reynolds number $\mathcal{R}$					& $22$				& $1$--$4$						\cr 
Mature ripple wavelength $\lambda$ (m)					& $0.14$ (d)			& $4$--$6$ (e), $4.5$ (f), $2$--$4$ (g) 	\cr
\hline
\end{tabular*}

\vspace*{0.1cm}

{{\bf Table S1}. Characteristic values of relevant parameters for Earth and Mars.} \\
{\bf Data sources}:\\
(a) {Elbelrhiti H, Andreotti B, and Claudin P. (2008) Barchan dune corridors: field characterization and investigation of control parameters. {\it J. Geophys. Res: Earth Surface}, 113(F2).}\\
(b) {Claudin P. and Andreotti B. (2006) A scaling law for aeolian dunes on Mars, Venus, Earth, and for subaqueous ripples. {\it Earth Planet. Sci. Lett.} 252, 30.}\\
(c) {Kok J.F., Parteli E.J.R., Michaels T.I., Karam D.B. (2012) The physics of wind-blown sand and dust. {\it Rep. Prog. Phys.} 75: 106901.}\\
(d) {Andreotti B., Claudin P., Pouliquen O. (2006) Aeolian sand ripples: experimental evidence of fully developed states. {\it Phys. Rev. Lett.} 96: 02800).}\\
(e) {Silvestro S., Fenton L.K., Vaz D.A., Bridges N.T., Ori. G.G. (2010) Ripple migration and dune activity on Mars: Evidence for dynamic wind processes. {\it Geophys. Res. Lett.} 37: L20203.}\\
(f) {Bridges N et al. (2011) Planet-wide sand motion on Mars. {\it Geology} 40, 31.}\\
(g) {Bridges N.T. et al. (2012) Earth-like sand fluxes on Mars. {\it Nature} 485: 339. }
\end{table}

\end{document}